\documentclass[aip,reprint]{revtex4-1}
\usepackage{graphics}
\usepackage{graphicx,epsfig}
\usepackage{epstopdf}

\begin{document}


\title{A compact rotating dilution refrigerator} 



\author{M. J. Fear}
	\affiliation{School of Physics and Astronomy, The University of Manchester, Oxford Road, Manchester, M13 9PL, United Kingdom}

\author{P. M. Walmsley}	
	\affiliation{School of Physics and Astronomy, The University of Manchester, Oxford Road, Manchester, M13 9PL, United Kingdom}

\author{D. A. Chorlton}	
	\affiliation{School of Physics and Astronomy, The University of Manchester, Oxford Road, Manchester, M13 9PL, United Kingdom}

\author{D. E. Zmeev}\altaffiliation{Department of Physics, Lancaster University, Lancaster, LA1 4YB, United Kingdom} 	
	\affiliation{School of Physics and Astronomy, The University of Manchester, Oxford Road, Manchester, M13 9PL, United Kingdom}
	
\author{S. J. Gillott}	
	\affiliation{School of Physics and Astronomy, The University of Manchester, Oxford Road, Manchester, M13 9PL, United Kingdom}
	
\author{M. C. Sellers}	
	\affiliation{School of Physics and Astronomy, The University of Manchester, Oxford Road, Manchester, M13 9PL, United Kingdom}

\author{P. P. Richardson}	
	\affiliation{School of Physics and Astronomy, The University of Manchester, Oxford Road, Manchester, M13 9PL, United Kingdom}

\author{H. Agrawal}\altaffiliation{Oak Ridge National Laboratory, Oak Ridge, TN 37831, USA.}	
	\affiliation{Oxford Instruments NanoScience, Tubney Woods, Abingdon, Oxfordshire, OX13 5QX, United Kingdom}

\author{G. Batey}	
	\affiliation{Oxford Instruments NanoScience, Tubney Woods, Abingdon, Oxfordshire, OX13 5QX, United Kingdom}

\author{A. I. Golov}	
	\affiliation{School of Physics and Astronomy, The University of Manchester, Oxford Road, Manchester, M13 9PL, United Kingdom}
	

\date{\today}

\begin{abstract}
We describe the design and performance of a new rotating dilution refrigerator that will primarily be used for investigating the dynamics of quantized vortices in superfluid $^4$He. All equipment required to operate the refrigerator and perform experimental measurements is mounted on two synchronously driven, but mechanically decoupled, rotating carousels. The design allows for relative simplicity of operation and maintenance and occupies a minimal amount of space in the laboratory. Only two connections between the laboratory and rotating frames are required for the transmission of electrical power and helium gas recovery. Measurements on the stability of rotation show that rotation is smooth to around $10^{-3}$\,rad\,s$^{-1}$ up to angular velocities in excess of 2.5\,rad\,s$^{-1}$. The behavior of a high-Q mechanical resonator during rapid changes in rotation has also been investigated.
\end{abstract}

\pacs{07.20.Mc 47.80.-v 47.32.Ef 47.37.+q}

\maketitle 

\section{Introduction}
Rotation has proven to be a vital tool for investigating the properties and dynamics of quantized vortices in superfluid $^4$He\,\cite{donnellybook}, the superfluid phases of $^3$He\,\cite{finne06} and Bose-Einstein condensates\cite{anderson10}. The equilibrium state of a bulk superfluid rotating with angular velocity, $\Omega$, is an array of rectilinear vortices with areal density $2\Omega/\kappa$, where $\kappa$ is the quantum of circulation ($\kappa=h/m=1.0\times10^{-3}$\,cm$^2$\,s$^{-1}$ for superfluid $^4$He). Rotation of a superfluid is thus the analogue of applying a magnetic field to a type II superconductor. Furthermore, dynamic changes in rotation have also proven to be useful: turbulent vortex tangles can be created by impulsive spin-down \cite{walmsley08,hosio12}; vortex waves \cite{walmsley12} and vortex sheets \cite{eltsov02} can occur when subject to an oscillatory component of rotation. The ability to rotate cryogenic apparatus can also be useful for some branches of astrophysics and particle physics\cite{oguri13}.

Studies of superfluid $^4$He under rotation now span over 60 years\cite{osborne50,hall60,andro67,donnelly56,reppy60}. There have been two approaches used that permit rotation and these are dependent on the temperatures required. At temperature above 1\,K, a rotating container of helium filled from a reservoir that remains stationary is relatively simple to implement and was the approach taken for many of the early studies of rotating helium \cite{hall60,andro67}. In the millikelvin regime, the general approach has been to rotate the entire cryostat due to the difficulty in maintaining thermal contact with an experimental container that rotates in isolation to the rest of the refrigerator. As a result of this increased complexity, only a small minority of dilution refrigerator cryostats are capable of rotation \cite{helsinki03,berkeley80,cornell81,helsinki83,helsinki89,berkeley90,manchester94,issp95,manchester04,issp03,riken06}. The majority of these instruments have also included a nuclear demagnetization stage for reaching the submillikelvin temperatures required for superfluid $^3$He which can further increase the bulk and sophistication of these cryostats. However, there are several reasons why a comparatively simple rotating cryostat that can cool samples of $^4$He to below 100\,mK would be a powerful tool. Our chief motivation is that the fundamentally important zero-temperature limit of the hydrodynamics of superfluid $^4$He occurs below 0.4\,K and rotation can be used to generate vortex arrays and turbulent tangles as well as allowing the calibration of vortex detection techniques. Furthermore, recent research on rotating solid $^4$He has produced intriguing and controversial results \cite{choi10,choi12,yagi11} that further highlight the importance of rotation in the millikelvin regime.

The primary requirement for a rotating cryostat is the need for smooth rotation with low vibration levels whilst maintaining an experiment at very low temperatures for the duration of the measurement. Any nuisance rotational or vibrational noise will produce unwanted excitation of vortex waves (such as Kelvin waves) due to the very low levels of dissipation at low temperatures which, at sufficient amplitudes, would result in reconnections between neighboring vortices and the destruction of the vortex array. Packard and co-workers\cite{berkeley80} failed to observe a vortex array in pure $^4$He at low temperatures, presumably due to excessive vibrational or rotational noise. Instead, they had to add a small amount of $^3$He to damp nuisance vortex motion so that images of vortex arrays could be obtained. This resulted in a widespread misconception that vortex arrays in pure $^4$He could not be obtained in the zero temperature limit due to the extremely small damping. More recently, Walmsley and Golov\cite{walmsley12} were able to create vortex arrays in a rotating container of isotopically pure $^4$He at 0.2\,K. They investigated the stability of the vortex array by adding a small oscillatory AC component of rotation to the steady value. The transition to turbulence in the bulk fluid was observed to occur for AC amplitudes as low as 0.03\,rad\,s$^{-1}$ at $\Omega=1.5$\,rad\,s$^{-1}$ indicating the importance of smooth steady rotation for the existence of vortex arrays.

All connections between the laboratory and rotating frames will produce some friction which can contribute to rotational noise, particularly if the level of friction varies with the rotational position of the cryostat (as happened in Ref.\,\onlinecite{manchester90}). So far, there have been two approaches to maintaining mixture circulation in the dilution refrigerator during rotation. One approach avoids the use of rotating vacuum seals by utilizing cryopumps to run the refrigerator in single-shot mode \cite{helsinki03,helsinki83,helsinki89}. The main disadvantage is the time consuming task of disconnecting and reconnecting the cryostat from the conventional vacuum pumps (which often occupy a separate room) in the laboratory frame at the start and end of each day to regenerate the cryopumps. Alternatively, other rotating cryostats use rotating vacuum seals, consisting of either lubricated rubber shaft seals\cite{berkeley80,berkeley90,manchester90,manchester94} or magnetic fluid based seals\cite{issp95,issp03,riken06}, to connect the stationary vacuum pumps to the cryostat which allows continuous operation regardless of whether the cryostat is stationary or rotating. This typically requires a complex assembly of up to four concentric rotating seals (for $^3$He-$^4$He mixture circulation, 1\,K pot pumping and $^4$He bath boil-off) and while generally reliable, any leak can lead to time-consuming repairs.

\section{Overview of design}

For the instrument described in this paper, we have taken the simplest possible approach and mounted all equipment required to run the refrigerator in the rotating frame. Advances in technology have resulted in compact and powerful vacuum pumps, with comparatively low levels of vibration, along with computer controlled gas-handling systems that are ideally suited for this application. Note that the inclusion of all equipment in the rotating frame limits the maximum angular velocity to $\simeq$3 rad\,s$^{-1}$ in contrast to some other cryostats which have used a different approach and as a result are capable of rotating above 25\,rad\,s$^{-1}$ (see Ref. \onlinecite{issp03} for further details). Our design philosophy from the outset has been to build a compact cryostat that is user-friendly and easy to maintain but without sacrificing the requirement for smooth rotation and low vibration levels. This approach rules out using ``dry" cryogen-free technology as this would involve a complex high pressure rotating seal in order to connect the rotating cryocooler to the bulky compressor in the laboratory frame (as used in Ref. \onlinecite{oguri13}). Also, our access to an in-house helium liquefier meant that a traditional ``wet" system was the most effective and appropriate solution.

\begin{figure*}[t]
\includegraphics[width=17.5cm]{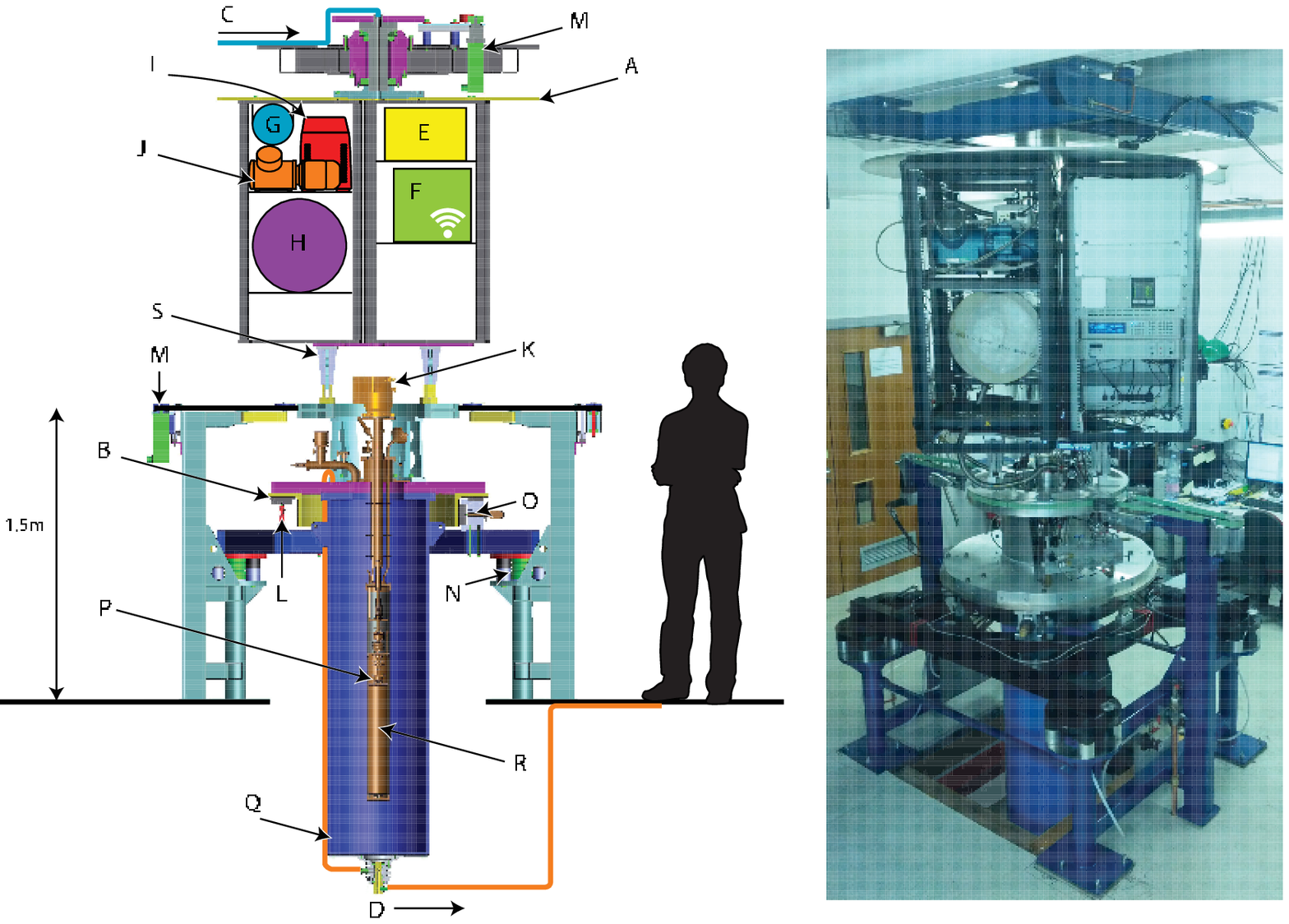}
\caption{General assembly diagram and photograph of the new rotating dilution refrigerator. The key features are labeled as follows: (A) upper carousel; (B) lower carousel; (C) single-phase electrical supply; (D) helium gas recovery via rotating seal; (E) gas handling manifold; (F) computer; (G) compressed air storage tank; (H) mixture storage tank; (I) forepump; (J) helium compressor; (K) turbo pump; (L) lift air pad mounted on a load cell; (M) drive motors; (N) air spring vibration isolators; (O) journal (centering) air pad; (P) mixing chamber of the dilution refrigerator; (Q) helium dewar; (R) available space for experimental cells and (S) overlapping aluminium arms to prevent misalignment of the two carousels.\label{fig_cryostat}}
\end{figure*}

The general assembly and an accompanying photograph of the cryostat is shown in Fig.\,\ref{fig_cryostat}. The apparatus occupies a minimal amount of space in our laboratory with a spatial footprint of only 3\,m$^2$ and a height of 4.4\,m. The key feature of the design is that there are two separate, mechanically decoupled, rotating carousels (as advocated in Ref. \onlinecite{helsinki03}). The cryogenic insert and helium dewar are mounted on the bottom carousel along with sensitive equipment, such as preamplifiers, for experimental measurements whereas all vibrationally noisy equipment, such as mechanical vacuum pumps and electronics with cooling fans, are attached to the upper carousel. The only flexible links between the two carousels are two gas lines, two pipes for cooling water and several electrical power and signal cables. As a consequence of our design, only two connections between the laboratory and rotating frames are required. The first is for $^4$He gas recovery at the bottom of the helium dewar, and the second is the single-phase electrical power supply that uses a Mercotac rotary electrical connector at the top of the upper carousel. All data communication between computers and instruments in the laboratory and rotating frames is achieved using standard commercial wireless LAN components.

The lower carousel consists of some components that were retrieved from a decommissioned rotating cryostat that was previously used in Manchester \cite{manchester90}. The cryogenic insert hangs from a 1m diameter aluminium disk that is mounted on top of an annular steel bearing ring. The steel ring is supported by three equally spaced 125\,mm diameter air pads that provide lift and a further three journal air pads that provide lateral support and define the central axis of the cryostat. The design of these aerostatic air bearings is similar to those used on some earlier cryostats\cite{berkeley90,manchester04,manchester90}. Each lift bearing has its own independently regulated air supply to allow the pressure to be tuned to suit its particular characteristics (typically $2.3\pm0.1$\,bar). The lift pads are each seated on a load cell with sufficient sensitivity to allow the carousel to be balanced to $\sim0.1$\,kg by adding small correction weights to the outer perimeter of the aluminium disk. One advantage of our design is that all the air bearings can be easily removed should the infrequent need for servicing arise. The lower carousel is supported by a horizontal steel frame which is vibrationally isolated from the lower steel support frame by four Firestone 25 airmount isolators. The vibration isolators are located at a diameter that is twice that of the bearing ring which helps to provide additional stability. Any motion of the horizontal steel frame is monitored by two orthogonal inclinometers.

The upper carousel consists of a 1.6\,m diameter aluminium disk that hangs from a steel support frame that is attached to the ceiling. The quality of rotation of this carousel is relatively unimportant and so conventional precision ball bearings are used. Furthermore, the robust design of the bearing support means that accurate balancing of the top carousel is not required. Three commercial steel $19^{\prime\prime}$ equipment racks, with 1.2\,m height, hang below the top aluminium plate. These racks house all equipment for running the dilution refrigerator as well as any electrical instruments required for experimental measurements that contain cooling fans or are not susceptible to the vibration levels present on this upper stage. With the present configuration, there is 20.5\,U (90\,cm) of $19^{\prime\prime}$ rack space that is available for housing any other equipment that may be required for a particular experiment.

Each carousel has its own independent drive provided by identical Aerotech BMS465 DC brushless, slotless servo motors. The slotless stator design eliminates cogging (periodic variation in the motor velocity due to slots in the motor windings) which is particularly important for smooth rotation at low angular velocities \cite{eltsov11}. The drive to the lower carousel is transferred using a flat drive belt with a pulley arrangement that ensures that no lateral force is applied but only a pure torque. The larger moment of inertia of the upper carousel requires a 5:1 planetary gearbox and toothed timing belt to apply the drive. The relative gearing of the two motors is governed electronically by two linked Aerotech Ensemble CL controllers. An infrared distance sensor continually monitors the position of the upper carousel relative to the lower one and dynamically adjusts the gain of the top motor to maintain the relative orientation. The possibility of the two carousels rotating more than $\pm6^{\circ}$ relative to each other (in the event of some malfunction) is prevented by two sets of overlapping aluminium arms that protrude from both carousels.

\section{Dilution refrigerator}

The dilution refrigerator is an Oxford Instrument Kelvinox MX-250 that has been specially modified so that the mixing chamber, heat exchangers, still and still pumping line are centered on the axis of rotation. A Joule-Thomson stage (JT) is used to assist with condensing incoming $^3$He gas instead of the more conventional 1\,K pot. This technique was used on some earlier rotating dilution refrigerators \cite{issp03}. In our case, this approach was taken primarily to simplify the gas handling system but also has the additional benefits of eliminating any vibration related to filling a 1\,K pot from the $^4$He bath and minimizes liquid $^4$He consumption.

Following precooling with cold $^4$He gas to around 20\,K and then filling the dewar with liquid $^4$He, the condensing in of the mixture and subsequent cooling of the mixing chamber to below 0.1\,K takes only 2.5\,hours. The base temperature of the refrigerator, with all experimental wiring in position is 12\,mK. The cooling power at 100\,mK is 290\,$\mu$W with a typical circulation rate of 450\,$\mu$mol\,s$^{-1}$. The mixture only requires 12\,litres (at STP) of expensive $^3$He gas. There is ample space (94.5\,mm diameter, 530\,mm height) for mounting several experiments below the mixing chamber. In order to increase the rigidity of the insert, a spoked ``wagon wheel"\cite{manchester04} using nylon covered drive cord is used to prevent relative motion between bottom of the 100\,mK radiation shield and the 4\,K vacuum can.

Almost all the components of the compact gas handling system are housed on the top carousel including the 10\,litre liquid nitrogen trap and 100\,litre mixture storage dump. In total, there are only six manual valves, which are required during the startup and shutdown of the refrigerator, and an additional five pneumatic computer-controlled valves. The compressed air required for actuating the pneumatic valves is supplied from a 5\,litre tank that is pressurized via a demountable connection to the laboratory air supply. The $^3$He-$^4$He mixture circulation path is completely oil-free. A Pfeiffer HiPace 300 turbo pump is directly mounted on the still pumping line on the central axis on the lower carousel. This location ensures that the pump can operate seamlessly regardless of whether the cryostat is rotating or not. A small 12\,V water pump and fan-cooled radiator (designed for cooling computer CPUs), mounted on the upper carousel, provides closed-circuit water cooling of the turbo pump. The two other pumps required for $^3$He-$^4$He mixture circulation are both mounted on the upper carousel: the turbo pump is backed by a multi-stage Roots forepump (Adixen ACP-28), and a KNF diaphragm compressor provides the $\simeq1$\,bar condensing pressure required at the JT stage during normal running of the refrigerator (a condensing pressure of $\simeq3$\,bar is required during the initial startup). Only 1.2\,kW of single-phase mains electrical power is required for running the entire refrigerator, including all diagnostic pressure and temperature measuring equipment. The largest diameter gas pipe in the system, and also the largest connection between the two carousels, is the NW-25 flexible hose that connects the back of the turbo pump to the forepump. All other hoses are 8\,mm o.d. and utilize double-ended shutoff quick connects which enables equipment to be easily removed from the system for servicing.

\section{Performance during rotation}
\begin{figure}[t]
\includegraphics[width=8.5cm]{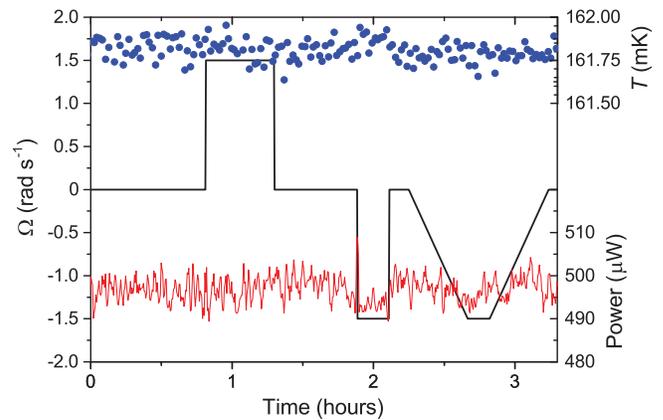}
\caption{Performance of the dilution refrigerator using closed-loop PID temperature stabilization during spin-up and spin-down of the cryostat. The central (black) curve shows angular velocity, $\Omega(t)$; the blue circles show the temperature of the mixing chamber (top right axis) and the bottom (red) curve shows the power applied to the heater mounted on the mixing chamber (bottom right axis). $\dot{\Omega}=10^{-1}$\,rad\,s$^{-2}$ for the first two rotations and $10^{-3}$\,rad\,s$^{-2}$ for the final rotation sequence. The mixing chamber thermometer is a calibrated ruthenium oxide resistor.\label{fig_rotation1}}
\end{figure}

One key requirement is that the performance characteristics (e.\,g. cooling power as a function of temperature) of a rotating dilution refrigerator should not depend on the angular velocity of rotation\cite{riken06} and ideally, there should be little noticeable change during sudden spin-up and spin-down of the cryostat. An example of the behavior of the new cryostat during a sequence of rotations is shown in Fig.\,\ref{fig_rotation1} with the PID temperature control of the mixing chamber set at a nominal 0.1618\,K. Neither the temperature nor the heat applied to the mixing chamber (and thus the cooling power) show any change with rotation within the scatter even during rapid changes of rotation with $\dot{\Omega}=\pm0.1$\,rad\,s$^{-2}$.

\begin{figure}[t]
\includegraphics[width=8.5cm]{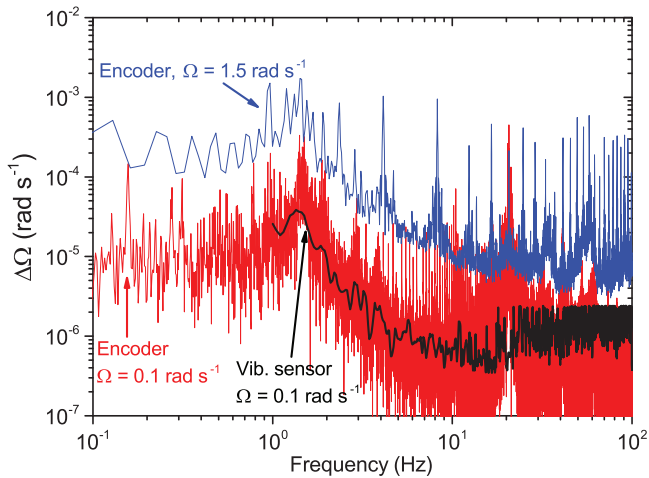}
\caption{Frequency spectrum of rotational noise measured using a rotary shaft encoder ($\Omega=0.1$ and 1.5\,rad\,s$^{-1}$) and horizontal vibration sensor ($\Omega=0.1$\,rad\,s$^{-1}$ only) mounted on the outer perimeter of the lower carousel. \label{fig_spectra}}
\end{figure}

\begin{figure}[t]
\includegraphics[width=8.5cm]{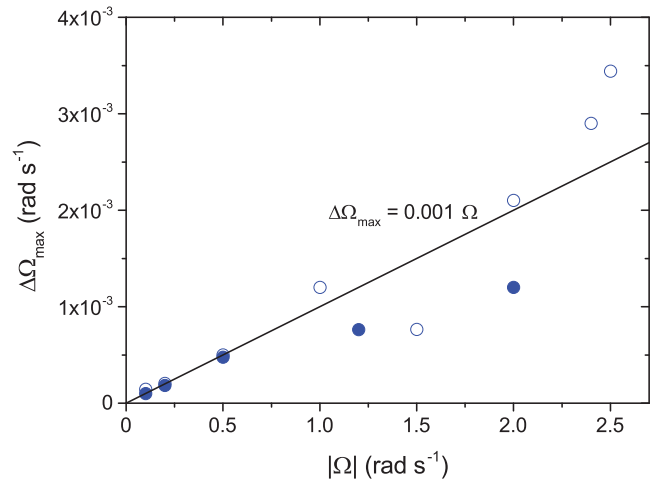}
\caption{Amplitude of rotational noise versus magnitude of angular velocity. The closed (open) symbols are for clockwise (anticlockwise) sense of rotation. The line indicates a linear fit with $\Delta\Omega_\mathrm{max}=0.001\Omega$. \label{fig_noise}}
\end{figure}

The rotational noise, $\Delta\Omega$, during steady rotation was monitored by measuring the velocity of the outer perimeter of the lower carousel using a rotary shaft encoder (with 1024 pulses per revolution resulting in 17,500 pulses per revolution of the cryostat). One potential complication of this technique is that additional noise due to the mechanical coupling of the encoder to the carousel can obscure the intrinsic noise due to any instabilities of rotation. Thus, a sensor (Geophone GS-11D with natural resonant frequency of 4.5z\,Hz) sensitive to vibration along a horizontal axis was used also mounted near the outer perimeter of the lower carousel. This sensor is suitable for use at low $\Omega\leq 0.1$\,rad\,$^{-1}$ but is unusable at higher $\Omega$ due to the resulting centrifugal force overloading the sensor. Some typical frequency spectra of $\Delta\Omega$ are shown in Fig.\,\ref{fig_spectra}. For $\Omega=0.1$\,rad\,s$^{-1}$, the two measurement techniques produce almost identical spectra although the spectra from the encoder are substantially more noisy and the encoder rotation frequency and harmonics are visible suggesting that some of the noise is generated by the encoder. $\Delta\Omega$ has a clear maximum at around 1.3\,Hz which is independent of $\Omega$. This is the natural resonant frequency of the lower carousel drive assembly for the particular belt tension used in this case. The smoothed value of $\Delta\Omega$ between 1 and 1.5\,Hz, $\Delta\Omega_\mathrm{max}$, as a function of $\Omega$ is shown in Fig.\,\ref{fig_noise}. A linear fit produces a slope $\Delta\Omega_\mathrm{max}/\Omega=0.001$ which is comparable to the rotational noise of other rotating dilution refrigerators \cite{helsinki03,riken06}. Note that adjusting the tension in the belt does change the frequency spectra of $\Delta\Omega$ and thus the tension can be changed, along with the servo parameters of the drive motor, to suit the requirements of a particular experiment but the values used here ensure that there is little slippage of the belt during rapid changes in $\Omega$ whilst also providing relatively stable steady rotation at constant $\Omega$. Another vibration sensor was used to monitor the vibration levels at the experimental stage below the mixing chamber during rotation. The maximum amplitude of vibration increased from 0.2\,nm with the cryostat stationary to 2\,nm during fast rotation. It should be noted that no increase in vibration due to the operation of the turbo pump on the lower carousel or the other pumps on the upper carousel was observed.

\begin{figure}[t]
\includegraphics[width=8.5cm]{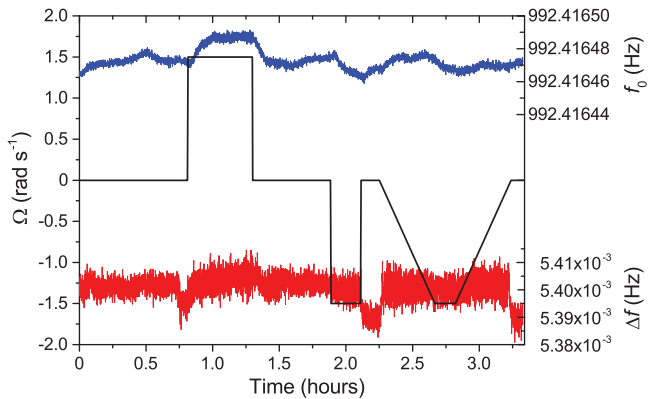}
\caption{Resonant frequency, $f_0$, and bandwidth, $\Delta f$, versus time for an empty torsional oscillator during spin-up and spin-down of the cryostat. The central (black) curve shows $\Omega(t)$ (left axis), the top (blue) curve shows $f_0$ (top right axis) and the bottom (red) curve shows $\Delta f$ (bottom right axis).\label{fig_TO}}
\end{figure}

A further test has utilized a torsional oscillator, a type of mechanical resonator that is widely used to probe various properties of quantum fluids and solids. Torsional oscillators have a very high quality factor at low temperatures and as a result they can be particularly sensitive to any vibrations resulting from rotation\cite{yagi10}. It is important that the frequency and width of the resonance of an oscillator are stable during rotation to allow both inertial and dissipative effects to be investigated. We have tested a torsional oscillator, intended for probing the dissipation due to Kelvin waves on quantized vortices in superfluid helium, on the new cryostat. The oscillator had a BeCu torsion stem with a Stycast epoxy torsion head containing a disk-shaped cavity. The oscillator has a resonant frequency of $f_0=992.4$\,Hz and a bandwidth (full width at half maximum) of $\Delta f=5.4$\,mHz, giving a $Q=f_0/\Delta f=1.8\times 10^5$. The motion of the oscillator was driven and detected capacitively with a typical amplitude of $\simeq 1$ nm (at the 15\,mm o.d. of the torsion head). The time-dependence of $f_0$ and $\Delta f$ during a sequence of rotations is shown in Fig.\,\ref{fig_TO}. There are occasional shifts in both $f_0$ and $\Delta f$ of around 10\,$\mu$Hz upon starting and stopping rotation although these are comparable to shifts that sometimes occurred when the cryostat was stationary. Following these shifts, the oscillator settles with a time-constant $\sim\,0.1$\,hours, which is comparable to the ring-down time during the free decay of oscillations. The stability of the oscillator during steady continuous rotation was comparable to when the cryostat was stationary. Furthermore, the observed shifts are tiny compared to those typically (often $\gg1$\,mHz) observed in superfluid and solid helium experiments and highlight the suitability of this cryostat for pursuing future research on these topics.

\section{Summary}
We have presented the design and performance of a rotating dilution refrigerator that is both compact and robust. The design is relatively simple since all equipment for running the refrigerator is located on the two rotating carousels which eliminates the need for complicated rotating vacuum seals and an extensive gas-handling system. The smoothness of rotation is comparable to other contemporary state-of-the-art rotating dilution refrigerators. The instrument is now fully operational and is currently being used for investigating the dynamics of vortices and quantum turbulence in superfluid $^4$He in the zero-temperature limit.

\begin{acknowledgments}
We are grateful to the staff of the mechanical workshop in Manchester for the skilled machining and fabrication of many of the components of this cryostat. We acknowledge many useful discussions with the ROTA group at Aalto University. This project was funded by an EPSRC Career Acceleration Fellowship awarded to P.M.W. (grant no. EP/I003738).
\end{acknowledgments}

\end{document}